\documentclass[preprint,review,12pt]{elsarticle}
\usepackage{graphicx}
\usepackage{amssymb}
 
\usepackage{amsmath}
\usepackage{dsfont}
\usepackage{bm}

\def\C{{\mathds C}}

\newcommand{\be}{\begin{equation}}
\newcommand{\ee}{\end{equation}}

\newcommand{\bzero}{{\mbox{\boldmath $0$}}}
\newcommand{\bI}{{\mbox{\boldmath $I$}}}
\newcommand{\bz}{{\mbox{\boldmath $z$}}}
\newcommand{\bn}{{\mbox{\boldmath $n$}}}
\newcommand{\bmm}{{\mbox{\boldmath $m$}}}

\newcommand{\bv}{{\mbox{\boldmath $v$}}}
\newcommand{\bvtilde}{{\mbox{\boldmath $\tilde{v}$}}}

\newcommand{\bp}{{\mbox{\boldmath $p$}}}

\newcommand{\bor}{{\mbox{\boldmath $r$}}}
\newcommand{\bee}{{\mbox{\boldmath $e$}}}

\newcommand{\bC}{{\mbox{\boldmath $C$}}}
\newcommand{\bA}{{\mbox{\boldmath $A$}}}
\newcommand{\bB}{{\mbox{\boldmath $B$}}}

\newcommand{\bU}{{\mbox{\boldmath $U$}}}

\newcommand{\bZ}{{\mbox{\boldmath $Z$}}}

\newcommand{\bS}{{\mbox{\boldmath $S$}}}
\newcommand{\bP}{{\mbox{\boldmath $P$}}}

\newcommand{\bSigma}{{\mbox{\boldmath $\Sigma$}}}
\newcommand{\balpha}{{\mbox{\boldmath $\alpha$}}}

\newcommand{\Pvtildeperp}{{\mbox{\boldmath $\bP^\perp_{\tilde{\bv}}$}}}
\newcommand{\Pvtilde}{{\mbox{\boldmath $\bP_{\!\tilde{\bv}}$}}}

\newproof{pf}{Proof}
\newproof{pf1}{Proof of proposition~\ref{Propo_KellyForsyth}}
\newtheorem{theorem}{Theorem}

\newcommand{\test}{\mbox{$
\begin{array}{c}
\stackrel{ \stackrel{\textstyle H_1}{\textstyle >} }{ 
\stackrel{\textstyle <}{ \textstyle  H_0} }
\end{array}
$}}

\begin{document}

\title{A novel approach to robust radar detection of range-spread targets}


\author[a]{Angelo Coluccia}
\ead{angelo.coluccia@unisalento.it}

\author[a]{Alessio Fascista}
\ead{alessio.fascista@unisalento.it}

\author[a]{Giuseppe Ricci 
\corref{cor1}}
\ead{giuseppe.ricci@unisalento.it}

\cortext[cor1]{Corresponding author}

\address[a]{DII, Universit\`a del Salento, Via Monteroni, 73100 Lecce, Italy.}

\begin{abstract}
This paper proposes a novel approach to robust radar detection of
range-spread targets embedded in Gaussian noise with unknown covariance matrix. The idea is to model the useful target echo in each range cell as the sum of a coherent signal plus a random component that makes the signal-plus-noise hypothesis more plausible in presence of mismatches. Moreover, an unknown power of the random components, to be estimated from the observables, is inserted to optimize the performance when the mismatch is absent. The generalized likelihood ratio test (GLRT) for the problem at hand is considered. In addition, a new parametric detector that encompasses the GLRT as a special case is also introduced and assessed. The performance assessment shows the effectiveness of the idea also in comparison to natural competitors. 
\end{abstract}

\begin{keyword}
Radar detection \sep distributed targets \sep GLRT \sep CFAR \sep robust detectors \sep mismatched signals
\end{keyword}

\maketitle

\section{Introduction}

The well-known problem of detecting the possible presence of a range-spread (or multiband) target is classically formulated as the following
hypothesis testing problem \cite{Kelly_techrep,CAI-WANG,Conte-DeMaio-Ricci,PartialObs}
\begin{equation}
\left\{
\begin{array}{lll}
H_{0}: & \bz_k =  \bn_{k}^{\text{\tiny P}}, & k=1, \ldots, K_P  \\
 &  \bor_k = \bn_{k}^{\text{\tiny S}}, & k=1, \ldots, K_S  \\ 
H_{1}: & \bz_k = \alpha_k \bv + \bn_{k}^{\text{\tiny P}}, & k=1, \ldots, K_P \\
 &  \bor_k = \bn_{k}^{\text{\tiny S}}, & k=1, \ldots, K_S  \\ 
\end{array} 
\right.
\label{eq:classical_hyp_test}
\end{equation}
where
the $\bz_k \in \C^{N \times 1}$, $k=1, \ldots, K_P$,
and the $\bor_k \in \C^{N \times 1}$, $k=1, \ldots, K_S$,
denote 
returns from primary data (i.e., cells under test) and secondary (or training) data, respectively.
As to $N$, it is the number of processed samples from each range cell; it might be the number of antenna array elements times the number of pulses.
Under the signal-plus-noise hypothesis $H_1$
the cells under test contain coherent returns from the target; namely
a known vector, say $\bv \in \C^{N \times 1}$, up to a complex factor, say $\alpha_k$, different
from one cell to another.
Moreover, the noise terms $\bn_{k}^{\text{\tiny P}}$, $k=1, \ldots, K_P$, and $\bn_{k}^{\text{\tiny S}}$, $k=1, \ldots, K_S,$
are modeled as independent and identically distributed random vectors ruled by the complex normal distribution
with zero mean and unknown (Hermitian) positive definite matrix $\bC$;  in symbols, 
we write (for the marginal distribution) $\bn_{k}^{\text{\tiny P}},\bn_{k}^{\text{\tiny S}} \sim {\cal CN}_N (\bzero, \bC)$. 
Modeling the $\alpha_k$s as unknown deterministic parameters
returns a complex normal distribution for the $\bz_k$ under both hypotheses; the non-zero mean of 
the received vector under $H_1$ makes it possible to discriminate between the two hypotheses, 
using the generalized likelihood ratio test (GLRT) and the ad hoc procedure also known as two-step GLRT-based design procedure \cite{Conte-DeMaio-Ricci}. 
A more general framework for multidimensional/multichannel signal detection in homogeneous Gaussian disturbance (with unknown covariance matrix and unknown structured deterministic interference) is considered in \cite{Ciunzo}.
Detection of distributed (or multiband) targets has also been addressed in presence of compound-Gaussian noise,  using for instance Rao and Wald tests \cite{Lombardo,ConteDeMaio,Guan2011}.

Many  works have addressed the problem of enhancing either the selectivity or the robustness of  adaptive detectors to  mismatches. In fact, 
a selective detector is desirable for accurate target localization. Instead, when a radar is working in searching mode, a certain level of robustness to mismatches is preferable. 
More generally, signal mismatches may occur due to miscalibration in the array, uncertainties about the target's angle of arrival or the Doppler frequency, etc. \cite{BOR-MorganClaypool}.

In particular, the cone idea has been used in \cite{Bandiera-Orlando-Ricci-conidistribuiti} to design robust detectors.
To increase instead the selectivity, 
the hypothesis testing problem (\ref{eq:classical_hyp_test}) can be modified, similarly to the 
adaptive beamformer orthogonal rejection test
(ABORT) formulation \cite{Pulsone},
introducing fictitious signals under the null
hypothesis; in \cite{BBR_WA} such signals are supposed orthogonal to the nominal steering vector in the 
whitened observation space.
In \cite{Orlando} a modification of the ABORT idea  is also proposed to come up with selective detectors for distributed targets in homogeneous or partially-homogeneous environments. 

The useful signal can also be modeled as a random vector that 
modifies the covariance matrix of the noise component \cite{Friedlander,Ricci}. In particular, a known steering vector multiplied by a complex normal random variable, independent of the (complex normal) noise term, produces  
a rank one modification of the noise covariance matrix. Interesting properties  in terms of either rejection capabilities or robustness to mismatches on the nominal steering vector can be obtained by considering this model, depending on the way the test is solved and possibly on the presence of a fictitious signal under the null hypothesis \cite{CAMSAP,Besson_collaboration}. 

More recently, a framework to design robust decision schemes  for point-like targets has been proposed \cite{CISS2018,BCRSigProc2018}.
The idea is to add to the $H_1$ hypothesis a random component that makes it more plausible, hence hopefully the detector more robust to mismatches on the nominal steering vector $\bv$. Such a random component is modeled as a zero-mean Gaussian vector with covariance matrix $\nu \bSigma$, where $\nu$ is an unknown, nonnegative factor.
In case of mismatch,  $\bSigma$ captures part of the signal leakage and the detector is more inclined to decide for $H_1$. For matched signals, instead,
the unknown factor $\nu$ limits the loss with respect to the GLRT that assumes $\nu=0$ (i.e., Kelly's detector).
In particular, when $\bm{\Sigma} = \bC$ is chosen, the 
GLRT (for point-like targets) 
is more robust than existing receivers that guarantee no loss under matched conditions with respect to Kelly's detector. Moreover, it has the constant false alarm rate (CFAR) property and a computational complexity comparable to that of Kelly's detector. It also lends itself to a parametric detector whose robustness can be controlled by a tunable parameter. Finally,
detection probabilities ($P_d$s)  of the GLRT and, more generally, of the parametric detector, depend only on the actual signal-to-noise ratio and the cosine squared between the whitened versions of the actual and the nominal steering vectors, say $\cos^2 \theta$. 
Thus, although it might be argued that the choice
$\bm{\Sigma} = \bC$ does not have a physical meaning, it leads to desirable behaviors under matched and mismatched conditions and, in particular, it guarantees performance depending on a meaningful measure of the mismatch (namely on $\cos^2 \theta$).

Motivated by the above results,  we investigate the potential of this approach
for robust detection of range-spread targets,
i.e., we extend the modeling idea in \cite{CISS2018,BCRSigProc2018} to the case of a target occupying more than one cell in range. In particular, we derive a robust GLRT-based detector for range-spread targets and also propose an ad hoc parametric receiver to obtain additional flexibility in the level of robustness. We also prove that such detectors have the CFAR property and that their $P_d$ depends on  the target amplitudes $\alpha_k$ only through the corresponding energy. The analysis confirms the validity of the considered approach also in comparison to natural competitors.

The paper is organized as follows: next section is devoted to the derivation of the proposed detectors; 
Section~III addresses their analysis
also in comparison to natural competitors by Monte Carlo simulation. Concluding remarks are given in Section~IV.

\section{Derivation of the GLRT and of the parametric detector for distributed targets}

Let us consider the following binary hypotheses testing problem
$$
\left\{
\begin{array}{lll}
H_0:  & \bz_k \sim {\cal CN}_N(0, \bC), & k=1, \ldots, K_P \\
& \bor_{k} \sim {\cal CN}_N(0, \bC), & k=1, \ldots, K_S  \\
H_1:  & \bz_k \sim {\cal CN}_N(\alpha_k  \bv, (1+\nu) \bC), & k=1, \ldots, K_P \\
& \bor_k \sim {\cal CN}_N(0, \bC), & k=1, \ldots, K_S
\end{array}
\right.
$$
where 
the positive definite matrix $\bC$, $\nu \geq 0$, and $\alpha_k \in \C$
are unknown quantities while $\bv \in \C^{N \times 1}$ is a known vector. Notice that the random components, introduced to make the $H_1$ hypothesis more plausible in presence of mismatches, give rise to the term $\nu \bC$ in the covariance matrix of the $\bz_k$s (which is present only under $H_1$).
Moreover, suppose that $\bz_1, \ldots, \bz_{K_P}, \bor_1, \ldots, \bor_{K_S}$ 
are independent random vectors under both hypotheses.
Finally, assume that $K_S \geq N$.

For future reference define
$\bZ=[\bz_1\ \cdots\ \bz_{K_P}]$,
$\bZ_{\alpha}=[\bz_1-\alpha_1 \bv\ \cdots\ \bz_{K_P}-\alpha_{K_P} \bv]$,
and $\balpha=[\alpha_1 \ \cdots \ \alpha_{K_P}]^T$ with $^T$ denoting, in turn, the transpose operator.
The corresponding GLRT  is given by
\be
\Lambda(\bZ, \bS)=
\frac{
\displaystyle{
\max_{\bC >0} \max_{\nu \geq 0} \max_{\balpha \in \C^{K_P}} f( \bZ, \bS | \bC, \nu, 
\balpha, H_1)
}}{
\displaystyle{\max_{\bC >0} f(\bZ, \bS | \bC, H_0)
}}
\test \eta
\label{eq:1S-GLRT_D}
\ee
where 
\begin{eqnarray*}
f( \bZ, \bS | \bC, \nu, \balpha, H_1) &=&
 \frac{c}{(1+\nu)^{NK_P} \det^{K_P+K_S}(\bC)} \\ &\times&
e^{- {\displaystyle \mathrm{tr}} \left\{ \bC^{-1} \left[  
\frac{1}{1+\nu} \bZ_{\alpha}
\bZ_{\alpha}^{\dagger}
 +  \bS \right] \right\} }
\end{eqnarray*}
and
\begin{eqnarray*}
f( \bZ, \bS | \bC, H_0) &=&
\frac{c}{\det^{K_P+K_S}(\bC)} \\ &\times&
e^{- {\displaystyle \mathrm{tr}} \left\{ \bC^{-1} \left[  
\bZ \bZ^{\dagger}
 +  \bS \right] \right\} }
 \end{eqnarray*}
denote the joint probability density functions (PDFs) of $\bz_1, \ldots, \bz_{K_P},\bor_1, \ldots, \bor_{K_S}$
under $H_1$ and $H_0$, respectively, with $\bS$ defined as $K_S$ times the sample covariance matrix based on training data,
i.e.,
\be
\bS= \sum_{k=1}^{K_S} \bor_k \bor_k^{\dagger}
\ee
and $c=\frac{1}{\pi^{N(K_P+K_S)}}$.
As to $^{\dagger}$, it denotes the Hermitian (i.e., conjugate transpose) operator, while $\det$, ${\displaystyle \mathrm{tr}}$, and $(\cdot)^{-1}$ are the determinant, the trace, and the inverse of the non-singular matrix argument, respectively.
Finally, $\eta$ is the detection threshold to be set according to the desired probability of false alarm ($P_{fa}$).

Maximization over $\bC$ can be performed as in \cite{Kelly};
we have that
$$
\widehat{\bC}_0
=
\frac{1}{K_P+K_S} \left[ \bZ \bZ^{\dagger} + \bS \right]
$$
and
$$
\widehat{\bC}_1(\nu, \balpha)
=
\frac{1}{K_P+K_S} \left[ \frac{1}{1+\nu} \bZ_{\alpha} \bZ_{\alpha}^{\dagger} + \bS \right].
$$
Thus, plugging the above expressions for $\bC$ into equation~(\ref{eq:1S-GLRT_D}), after some algebra,
yields
\begin{eqnarray*}
\Lambda^{\prime}(\bZ, \bS) &=& \Lambda^{\frac{1}{K_P+K_S}}(\bZ, \bS) 
\\ &=&
\frac{
\det \left(  \bZ \bZ^{\dagger} + \bS \right)
}{
\displaystyle{\min_{\nu \geq 0, \balpha \in \C^{K_P}}}
(1+\nu)^{\frac{NK_P}{K_P+K_S}}  \det \left(\frac{
\bZ_{\alpha} \bZ_{\alpha}^{\dagger}}{\nu+1} + \bS \right) 
}
\label{eq:1S-GLRT_D-1}
\end{eqnarray*}
Minimization with respect to $\balpha$ (i.e., $\alpha_1, \ldots, \alpha_{K_P}$) can be conducted using the following proposition that makes use of the fact that $\bS$ is positive definite (with probability one); in fact, $K_S \geq N$ and $\bC>0$.
\begin{theorem}
\label{Propo_KellyForsyth}
Suppose that $\bS$ is a positive definite matrix.
It follows that
\begin{eqnarray*}
m_{\min} &=& \min_{\balpha \in \C^{K_P}} 
\det \left(\frac{
\bZ_{\alpha} \bZ_{\alpha}^{\dagger}}{\nu+1}  + \bS \right)
\\ &=&
\det (\bS) \det \left[ \frac{\Pvtildeperp \bS^{-1/2} \bZ \bZ^{\dagger} \bS^{-1/2} \Pvtildeperp}{
\nu+1}  + \bI_{N} \right]
\\ &=&
\det \left[ \frac{1}{\nu+1} \bZ_{\widehat{\alpha}}  \bZ^{\dagger} _{\widehat{\alpha}}
+ \bS \right]
\end{eqnarray*}
with
$$
\bZ_{\widehat{\alpha}}= \left[ \bz_{1} - \widehat{\alpha}_1  \bv \cdots \bz_{K_P} - \widehat{\alpha}_{K_P} \bv  \right]
$$
where $\tilde{\bv}=\bS^{-1/2}\bv$
and
the $\widehat{\alpha}_k$s, given by
$$
\widehat{\alpha}_k = \frac{  \bv^{\dagger} \bS^{-1} \bz_{k} }{\bv^{\dagger} \bS^{-1} \bv},
\quad k=1, \ldots, K_P,
$$
are the coordinates of the minimizer of the function under study. Finally,
$\Pvtilde$ and $\Pvtildeperp = \bI_N-\Pvtilde$ are the projection matrices onto 
the space spanned by $\bvtilde$ and its orthogonal complement, respectively.
\end{theorem}

\begin{pf}
See Appendix A.
\hfill$\square$
\end{pf}

Notice that this result was firstly derived in \cite{Kelly_techrep} in a more general framework. We have given here a new and more compact proof of the result by focusing on the form we are interested in.
Using it we can re-write the GLRT statistic as
\be
\frac{
\det \left(  \bS^{-1/2} \bZ \bZ^{\dagger} \bS^{-1/2} + \bI_{N} \right)
}{
\displaystyle{\min_{\nu \geq 0}}
(1+\nu)^{\frac{NK_P}{K_P+K_S}}  
\det \left[ \frac{\Pvtildeperp \bS^{-1/2} \bZ \bZ^{\dagger} \bS^{-1/2} \Pvtildeperp}{\nu+1} + \bI_{N} \right]
}
\label{eq:1S-GLRT_D-2}
\ee
or
\be
\frac{
\det \left( \bZ \bZ^{\dagger}  + \bS \right)
}{
\min_{\nu \geq 0}
(1+\nu)^{\frac{NK_P}{K_P+K_S}}  
\det \left[ \frac{1}{\nu+1}   \bZ_{\widehat{\alpha}} \bZ_{\widehat{\alpha}} ^{\dagger} + \bS \right]
}.
\label{eq:1S-GLRT_D-3}
\ee
Minimization of the denominator of equation~(\ref{eq:1S-GLRT_D-2})
(or equation~(\ref{eq:1S-GLRT_D-3}))
can be conducted resorting to the following proposition.

\begin{theorem}
Let $\bA \in \C^{N \times N}$ be a Hermitian positive semidefinite
matrix of rank $r$,  $1 \leq r \le N$, with non-zero eigenvalues $\lambda_1, \ldots, \lambda_r$.
The function
\begin{eqnarray}
\nonumber
f(\nu) &=& (1+\nu)^m \det  \left[ \frac{1}{\nu+1} \bA + \bI_N \right]
\\ &=&
(1+\nu)^m \prod_{i=1}^r \left( \frac{\lambda_i}{\nu+1} +1 \right),
\label{eq:function_lemma4}
\end{eqnarray}
with $m >0$,
attains its minimum with respect to $\nu \in [0, +\infty)$ at
\be
\widehat{\nu}=
\left\{
\begin{array}{ll}
0, & \mbox{if } \sum_{i=1}^r \frac{\lambda_i}{\lambda_i+1}  \leq m \\
\overline{\nu}, & \mbox{otherwise}
\end{array}
\right.
\label{eq:nu_hat}
\ee
where $\overline{\nu}$ is the unique solution over $(0, +\infty)$
of the equation
\be
g(\nu)=\sum_{i=1}^r \frac{\lambda_i}{\lambda_i +1 + \nu}  = m
\label{eq:lemma4}
\ee
under the condition $\sum_{i=1}^r \frac{\lambda_i}{\lambda_i+1}  > m$.
\end{theorem}

\begin{pf}
First observe that
the function
$$
g(\nu)= \sum_{i=1}^r \frac{\lambda_i}{\lambda_i +1 + \nu}
$$
is strictly decreasing over $\nu \in [0, +\infty)$. In fact,
its derivative, given by
$$
\frac{\mathrm{d} g(\nu)}{\mathrm{d}\nu}= -\sum_{i=1}^r \frac{\lambda_i}{(\lambda_i +1 + \nu)^2},
$$
is negative over $\nu \in [0, +\infty)$. In addition, $\lim_{\nu \rightarrow +\infty} g(\nu)=0$;
thus, if $g(0)=\sum_{i=1}^r \frac{\lambda_i}{\lambda_i+1}  > m$,
equation~(\ref{eq:lemma4}) admits a unique solution for $\nu >0$.

Moreover, 
$$
\lim_{\nu \rightarrow +\infty} f(\nu) = +\infty
$$
and
the derivative of function~(\ref{eq:function_lemma4}) is given by
\begin{eqnarray*}
\frac{\mathrm{d} f(\nu)}{\mathrm{d}\nu} &=&
m(1+\nu)^{m-1} \prod_{i=1}^r \left( \frac{\lambda_i}{\nu+1} +1 \right)
\\ &-& (1+\nu)^m \sum_{i=1}^r \frac{\lambda_i}{(\nu+1)^2} \prod_{j=1, j\neq i}^r \left( \frac{\lambda_j}{\nu+1} +1 \right)
\\ &=&
(1+\nu)^{m-1} \prod_{i=1}^r \left( \frac{\lambda_i}{\nu+1} +1 \right)
\\ &\times&
\left[ m- 
\sum_{i=1}^r \frac{\lambda_i }{\lambda_i + \nu+1} \right]
\\ &=&
(1+\nu)^{m-1} \prod_{i=1}^r \left( \frac{\lambda_i}{\nu+1} +1 \right)
\left[ m- g(\nu) \right].
\end{eqnarray*}
Thus, if $\sum_{i=1}^r \frac{\lambda_i}{\lambda_i+1}  \leq m$, it follows that
$\frac{\mathrm{d} f(\nu)}{\mathrm{d}\nu} > 0$, $\forall \nu \in (0, +\infty)$ and $\frac{\mathrm{d} f(\nu)}{\mathrm{d}\nu}\big |_{\nu=0} \geq  0$ and, hence, 
the minimum of $f(\nu)$ is attained at $\nu=0$. 
Otherwise, the minimizer is the positive value of $\nu$ solving equation
$\frac{\mathrm{d} f(\nu)}{\mathrm{d}\nu} = 0$ or, equivalently, equation~(\ref{eq:lemma4}).
\hfill$\square$
\end{pf}

Using the above lemma with $m=\frac{NK_P}{K_P+K_S}$ and
\be
\bA= \Pvtildeperp \bS^{-1/2} \bZ \bZ^{\dagger} \bS^{-1/2} \Pvtildeperp
\label{eq:matrix_A_proposition2}
\ee
we can compute the GLRT that is 
\be
\frac{
\det \left(  \bS^{-1/2} \bZ \bZ^{\dagger} \bS^{-1/2} + \bI_{N} \right)
}{
(1+\widehat{\nu})^{\frac{NK_P}{K_P+K_S}}  
\det \left[ \frac{\Pvtildeperp \bS^{-1/2} \bZ \bZ^{\dagger} \bS^{-1/2} \Pvtildeperp}{\widehat{\nu}+1} + \bI_{N} \right]
}
\test \eta
\label{eq:The1S-GLRT_D}
\ee
with $\widehat{\nu}$ given by (\ref{eq:nu_hat}).
Equation (\ref{eq:The1S-GLRT_D}) can also be re-written as
\be
\frac{\det(\bZ \bZ^{\dagger} + \bS)}{(1 + \widehat{\nu})^{\frac{NK_P}{K_P+K_S}} \det \left[\frac{1}{1+\widehat{\nu}} \bZ_{\widehat{\alpha}}  \bZ_{\widehat{\alpha}} ^{\dagger} + \bS \right]}
\test \eta
\label{eq:The1S-GLRT_Dsecondform}
\ee
where $\bZ_{\widehat{\alpha}}$  is given in Theorem \ref{Propo_KellyForsyth}.
We notice that, under the condition
\be
\sum_{i=1}^r \frac{\lambda_i}{\lambda_i+1}  \leq m=\frac{NK_P}{K_P+K_S},
\label{eq:condition_nu0}
\ee
where the $\lambda_i$ are the non-zero eigenvalues
of the matrix $\bA$ in (\ref{eq:matrix_A_proposition2}),
the statistic of the GLRT
(\ref{eq:The1S-GLRT_D})
is equivalent to that of 
 the GLRT for homogeneous environment (equation (12) in \cite{Conte-DeMaio-Ricci}), i.e.,
\be
\frac{
\det \left(  \bS^{-1/2} \bZ \bZ^{\dagger} \bS^{-1/2} + \bI_{N} \right)
}{ 
\det \left[ \Pvtildeperp \bS^{-1/2} \bZ \bZ^{\dagger} \bS^{-1/2} \Pvtildeperp + \bI_{N} \right]
}
\test \eta
\label{eq:GLRT-H}
\ee
 and, in fact, $\widehat{\nu}=0$. Thus, the possible enhanced robustness of detector (\ref{eq:The1S-GLRT_D}) 
 with respect to
 the GLRT for homogeneous environment
can be ascribed to the use of the decision statistic corresponding to the condition complementary to (\ref{eq:condition_nu0}). 
As a consequence, it is also reasonable to investigate the behavior of a potentially even more robust detector obtained  by decreasing the probability to select the statistic (\ref{eq:GLRT-H}). In particular, we propose to   replace $m$ in (\ref{eq:condition_nu0}) with $m_{\epsilon}=\frac{NK_P}{K_P+K_S}\frac{1}{1+\epsilon}$, $\epsilon \geq 0$.
We also modify the decision statistic 
of the GLRT
(\ref{eq:The1S-GLRT_D}) by
replacing 
$m=\frac{NK_P}{K_P+K_S}$ with $m_{\epsilon}$.
Accordingly, we consider the following parametric detector
\be
\frac{
\det \left(  \bS^{-1/2} \bZ \bZ^{\dagger} \bS^{-1/2} + \bI_{N} \right)
}{
(1+\widehat{\nu}_{\epsilon})^{m_{\epsilon}}  
\det \left[ \frac{\Pvtildeperp \bS^{-1/2} \bZ \bZ^{\dagger} \bS^{-1/2} \Pvtildeperp}{\widehat{\nu}_{\epsilon}+1} + \bI_{N} \right]
}
\test \eta
\label{eq:parametric_detector}
\ee
with $\widehat{\nu}_{\epsilon}$ given by
\be
\widehat{\nu}_{\epsilon}=
\left\{
\begin{array}{ll}
0, & \mbox{if } \sum_{i=1}^r \frac{\lambda_i}{\lambda_i+1}  \leq m_{\epsilon} \\
\overline{\nu}_{\epsilon}, & \mbox{otherwise}
\end{array}
\right.
\label{eq:nu_hat_parametric_detector}
\ee
where $\overline{\nu}_{\epsilon}$ is the unique solution over $(0, +\infty)$
of the equation
\be
\sum_{i=1}^r \frac{\lambda_i}{\lambda_i +1 + \nu}  = m_{\epsilon}
\label{eq1:nu_hat_parametric_detector}
\ee
where the $\lambda_i$s are the non-zero eigenvalues 
of the matrix $\bA$ in (\ref{eq:matrix_A_proposition2}),
under the condition $\sum_{i=1}^r \frac{\lambda_i}{\lambda_i+1}  > m_{\epsilon}$. Notice that the parametric detector encompasses the GLRT as a special case for $\epsilon=0$.
\medskip

It is possible to prove that detectors (\ref{eq:The1S-GLRT_D}) and (\ref{eq:parametric_detector}) possess the CFAR property.
In fact, the following result holds true.
\begin{theorem}
The decision statistics (\ref{eq:The1S-GLRT_D})
and (\ref{eq:parametric_detector})
have a distribution independent of $\bC$ under the $H_0$ hypothesis.
\end{theorem}

\begin{pf}
It is obviously sufficient to prove the proposition focusing on the parametric detector.
First we highlight that the matrices
$$
\Pvtildeperp \bS^{-1/2} \bZ \bZ^{\dagger} \bS^{-1/2} \Pvtildeperp
$$
and $\bZ^{\dagger} \bS^{-1/2} \Pvtildeperp \bS^{-1/2} \bZ$
have the same non-zero eigenvalues \cite{Magnus}.
Then, observe that
\begin{eqnarray*}
\bZ^{\dagger} \bS^{-1/2} \Pvtildeperp \bS^{-1/2} \bZ &=& \bZ^{\dagger} \bS^{-1} \bZ -
\bZ^{\dagger} \bS^{-1/2} \Pvtilde \bS^{-1/2} \bZ
\\ &=& \bZ^{\dagger} \bC^{-1/2} \bC^{1/2} \bS^{-1} \bC^{1/2} \bC^{-1/2}\bZ
\\ &-& \bZ^{\dagger} \bC^{-1/2} \bC^{1/2} \bS^{-1} \bC^{1/2} \bC^{-1/2} \bv 
\\ &\times&
\!\!\!\!\! \left( \bv^{\dagger} \bC^{-1/2} \bC^{1/2} \bS^{-1} \bC^{1/2} \bC^{-1/2} \bv \right)^{-1} 
\\ &\times& \bv^{\dagger} \bC^{-1/2} \bC^{1/2} \bS^{-1} \bC^{1/2} \bC^{-1/2} \bZ
\\ &=& \bZ_0^{\dagger} \bS_0^{-1} \bZ_0
\\ &-& \bZ_0^{\dagger} \bS_0^{-1} \bv_0 
\left( \bv_0^{\dagger} \bS_0^{-1} \bv_0 \right)^{-1} \!\! \bv_0^{\dagger} \bS_0^{-1} \bZ_0
\\ &=& \bZ_0^{\dagger} \bU \bU^{\dagger} \bS_0^{-1} \bU \bU^{\dagger} \bZ_0
\\ &-& \bZ_0^{\dagger} \bU \bU^{\dagger} \bS_0^{-1}  \bU \bU^{\dagger} \bv_0 
\\ &\times& \left( \bv_0^{\dagger} \bU \bU^{\dagger} \bS_0^{-1} \bU \bU^{\dagger} \bv_0 \right)^{-1} 
\\ &\times& \bv_0^{\dagger} \bU \bU^{\dagger} \bS_0^{-1} \bU \bU^{\dagger} \bZ_0 
\end{eqnarray*}
where
$\bZ_0= \bC^{-1/2} \bZ$, $\bS_0^{-1} = \bC^{1/2} \bS^{-1} \bC^{1/2}$,
$\bv_0= \bC^{-1/2} \bv$, and $\bU^{\dagger}$ a unitary matrix that rotates $\bv_0$ onto the first vector 
$\bee_1=[1 \ 0 \cdots 0] \in \C^{N \times 1}$
of the canonical basis, i.e.,
$$
\bU^{\dagger} \bv_0=K \bee_1
$$
with $K$ a proper constant. 
It follows that
\begin{eqnarray*}
\bZ^{\dagger} \bS^{-1/2} \Pvtildeperp \bS^{-1/2} \bZ 
&=& \bZ_{u 0}^{\dagger} \bS_{u 0}^{-1}  \bZ_{u 0}
- \bZ_{u 0}^{\dagger} \bS_{u 0}^{-1} \bee_1 
\\ &\times& \left( \bee_1^{\dagger} \bS_{u 0}^{-1} \bee_1 \right)^{-1} 
\bee_1^{\dagger}  \bS_{u 0}^{-1}  \bZ_{u 0} 
\end{eqnarray*}
where $\bZ_{u 0}= \bU^{\dagger} \bZ_{0}$ and $\bS_{u 0}^{-1}= \bU^{\dagger} \bS_{0}^{-1} \bU$.

Finally, we observe that
\begin{itemize}
\item[1.]
$\widehat{\nu}_{\epsilon}$, given by equation~(\ref{eq:nu_hat_parametric_detector}), is a function of
$\bZ_{u 0}$ and $\bS_{u 0}^{-1}$ only;
\item[2.]
the determinant at the denominator of the decision statistic (\ref{eq:parametric_detector}) 
can be re-written as
\begin{eqnarray*}
&\det& \!\!\!\! \left[ \frac{1}{\widehat{\nu}_{\epsilon}+1}   \Pvtildeperp \bS^{-1/2} \bZ \bZ^{\dagger} \bS^{-1/2} \Pvtildeperp + \bI_{N} \right] \\ &=&
\det \left[ \frac{1}{\widehat{\nu}_{\epsilon}+1}    \bZ^{\dagger} \bS^{-1/2} \Pvtildeperp \bS^{-1/2} \bZ + \bI_{K_{P}} \right];
\end{eqnarray*}
\item[3.]
the numerator of the decision statistic (\ref
{eq:parametric_detector}) can be re-written as
\begin{eqnarray*}
\nonumber
&\det& \!\!\!\! \left[ \bS^{-1/2} \bZ \bZ^{\dagger} \bS^{-1/2} + \bI_{N} \right] 
\\ &=&
\det \left[ \bZ^{\dagger} \bS^{-1}   \bZ + \bI_{K_{P}} \right]
\\ &=&
\det \left[ \bZ_{u 0}^{\dagger} \bS_{u 0}^{-1}   \bZ_{u 0} + \bI_{K_{P}} \right].
\end{eqnarray*}
\end{itemize}
It turns out that the statistic of the parametric detector
is a function of $\bZ_{u 0}$ and $\bS_{u 0}^{-1}$ only.
Since $\bZ_{u 0}$ and $\bS_{u 0}^{-1}$ are independent random quantities and, in addition,
each of them
has a distribution independent of $\bC$, under $H_0$, 
it follows that
the statistic of the parametric detector
has a distribution independent of $\bC$ (under  $H_0$). 
\hfill$\square$
\end{pf}

It is also possible to prove that the $P_d$s of the proposed detectors
depend on the $\alpha_k$s only through $\sum_{k=1}^{K_P} | \alpha_k |^2$.
In fact, the following result holds true.
\begin{theorem}
The left-hand side of equations~(\ref{eq:The1S-GLRT_D})
and (\ref{eq:parametric_detector})
have distributions that, under $H_1$, depend on the $\alpha_k$s only through $\sum_{k=1}^{K_P} | \alpha_k |^2$.
\end{theorem}

\begin{pf}
The proof comes from observing that the decision statistics of the considered detectors depend on primary data only through the quantity
$\bZ \bZ^{\dagger}$. Since $\bz_1, \ldots, \bz_{K_P}$ are independent complex normal random vectors
and, in addition, under $H_1$, $\bz_i \sim {\cal CN}_N\left(\alpha_i \bp, (1+\nu)\bC \right)$,
with $\bp$ the possible mismatched steering vector ($\bp=\bv$ under matched conditions),
$\bZ \bZ^{\dagger}$ is a complex non-central Wishart distribution. It follows that the distribution of $\bZ \bZ^{\dagger}$
depends on  the $\bmm_k=\alpha_k \bp$s only through \cite{Tourneret,McKay}
$$
\sum_{k=1}^{K_P} \bmm_k \bmm_k^{\dagger}= \bp \bp^{\dagger} \sum_{k=1}^{K_P}
| \alpha_k |^2.
$$
\hfill$\square$
\end{pf}

\section{Performance analysis}

The analysis is conducted by Monte Carlo simulation.
The $P_{fa}$s and the $P_d$s are estimated through Monte Carlo counting techniques, based on $100/P_{fa}$ and $10^3$ independent trials, respectively. To limit the computational burden required for the threshold setting, we assume $P_{fa} = 10^{-4}$.

The noise components of the $\bm{z}_k$s and the $\bm{r}_k$s are generated as independent random vectors ruled by a zero-mean, complex normal distribution. As concerns the covariance matrix, we adopt as $\bm{C}$ the sum of a Gaussian-shaped clutter covariance matrix $\bm{R}_c$ plus a white (thermal) noise 10 dB weaker, that is, $\bm{C} = \bm{R}_c + \sigma^2_n \bm{I}_N$, where the $(i,j)$-th element of the matrix $\bm{R}_c$ is obtained as $[\bm{R}_c]_{i,j} =\mathrm{exp}\{-2\pi^2\sigma_f^2(i - j)^2\}$ and $\sigma_f = 0.05$  (which corresponds to a one-lag correlation coefficient of 0.95).
In addition, we assume a time steering vector, i.e.,
$\bm{v} = \left[1 \ e^{i 2\pi f_d} \ \cdots \ e^{i 2\pi(N-1)f_d} \right]^T$, 
with $N=16$ and a nominal value of the normalized Doppler frequency $f_d = 0.08$, a value chosen such that the target competes with the adopted lowpass spectrum of the disturbance (clutter plus thermal noise). 
The robustness of the proposed detectors is assessed by simulating a target with a mismatched signature $\bp$ having normalized Doppler frequency $f_d + \delta/N$ with $\delta=0.4$.
To quantify the mismatch between the assumed and the actual target steering vector, we define
\be 
\cos^2 \theta = \frac{|\bv^{\dagger} \bC^{-1} \bp|^2}{(\bv^{\dagger}\bC^{-1}\bv)(\bp^{\dagger}\bC^{-1}\bp)}.
\ee 
Thus, $\theta$ represents the mismatch angle between the nominal steering vector $\bv$ and its mismatched version $\bp$. Observe that $\cos^2 \theta = 1$ corresponds to perfect match while $\delta=0.4$ implies
$\cos^2 \theta = 0.46$.

The target amplitudes $\alpha_k, k=1,\ldots,K_P$, associated to the returning echoes, are generated deterministically according to the signal-to-noise ratio (SNR), defined as
\begin{equation}
\text{SNR} = \sum_{k=1}^{K_P} |\alpha_k|^2 \bp^{\dagger} \bm{C}^{-1}\bp.
\end{equation}
More precisely, since the performance of the detectors are independent of the specific values  of the $\alpha_k$s, in the following we assume that the total energy $\sum_{k=1}^{K_P} |\alpha_k|^2$ is equally distributed along the $K_P$ cells.

The proposed GLRT and the parametric detector are compared against the 
GLRT of equation (\ref{eq:GLRT-H}), see also \cite{Conte-DeMaio-Ricci}, and referred to 
in the following as GLRT-H.
Obvious references are also the generalized adaptive matched filter (GAMF) \cite{Conte-DeMaio-Ricci}
\begin{equation}
\sum_{k=1}^{K_P} \frac{|\bm{z}_k^{\dagger} \bm{S}^{-1} \bm{v}|^2}{\bm{v}^{\dagger}\bm{S}^{-1}\bm{v}} \test \eta_{\text{\tiny GAMF}}
\end{equation}
and the generalized adaptive subspace detector (GASD) \cite{Conte-DeMaio-Ricci}
\begin{equation}
\sum_{k=1}^{K_P} \frac{|\bm{z}_k^{\dagger} \bm{S}^{-1} \bm{v}|^2}{\bm{v}^{\dagger}\bm{S}^{-1}\bm{v} \sum_{h=1}^{K_P} \bm{z}_h^{\dagger}\bm{S}^{-1}\bm{z}_h} \test 
\eta_{\text{\tiny GASD}}.
\end{equation}

It is also worth observing that the GAMF and the GASD reduce to the well-known AMF \cite{Kelly-Nitzberg} and ACE, \cite{Asymptotically,ACE}, see also \cite{Gini}, respectively, for $K_P = 1$. As to the GLRT-H, it is a special case of the one proposed by Kelly and Forsythe \cite{Kelly_techrep} and reduces to Kelly's detector \cite{Kelly} for $K_P = 1$. 

\begin{figure}
	\centering
	\includegraphics[width=0.8\textwidth]{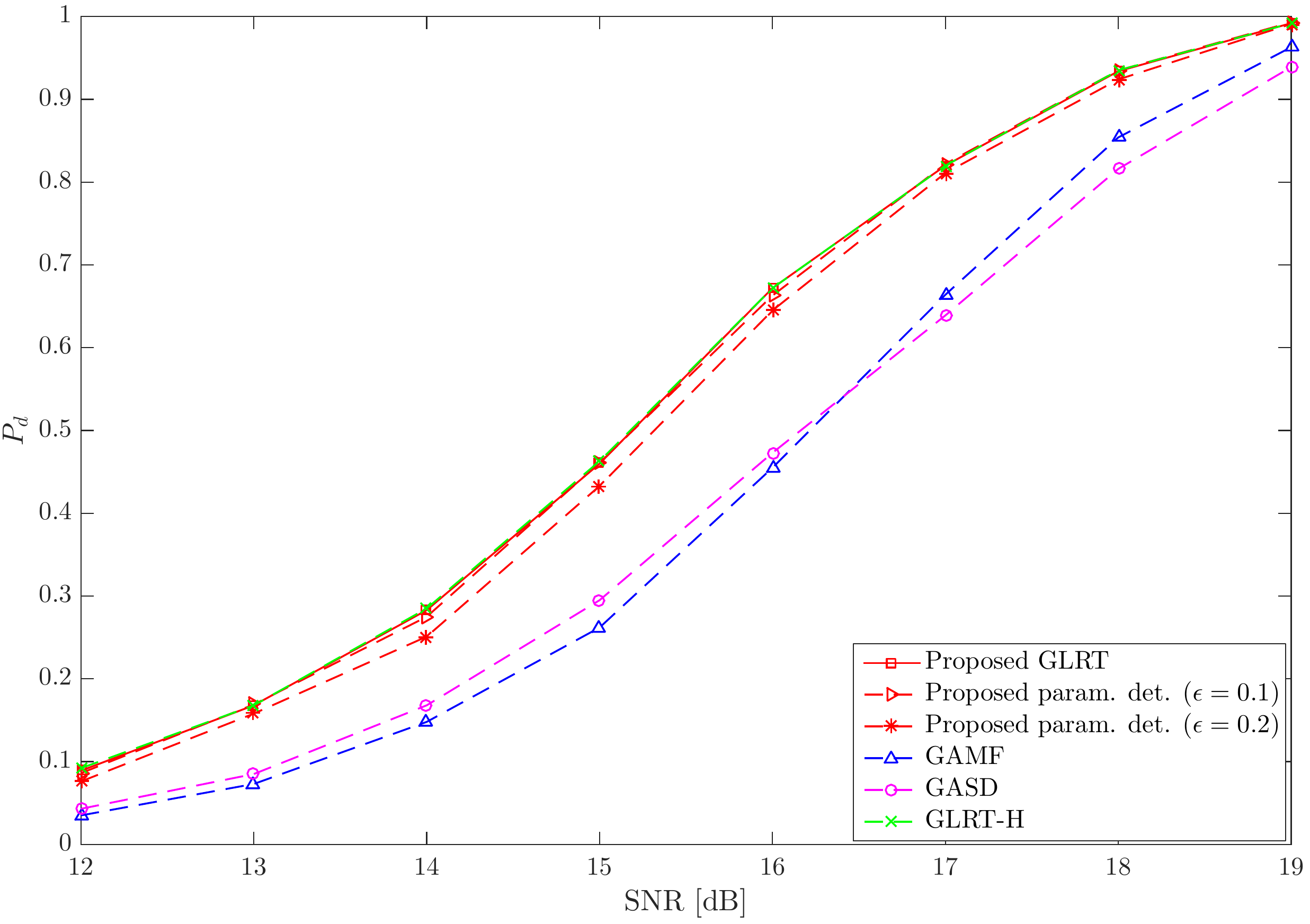}
	\caption{$P_d$ vs SNR under matched conditions, $N=16$, $K_P=4$, and $K_S=32$.}
	\label{fig:matchedKs32}
\end{figure}

\begin{figure}
	\centering
	\includegraphics[width=0.8\textwidth]{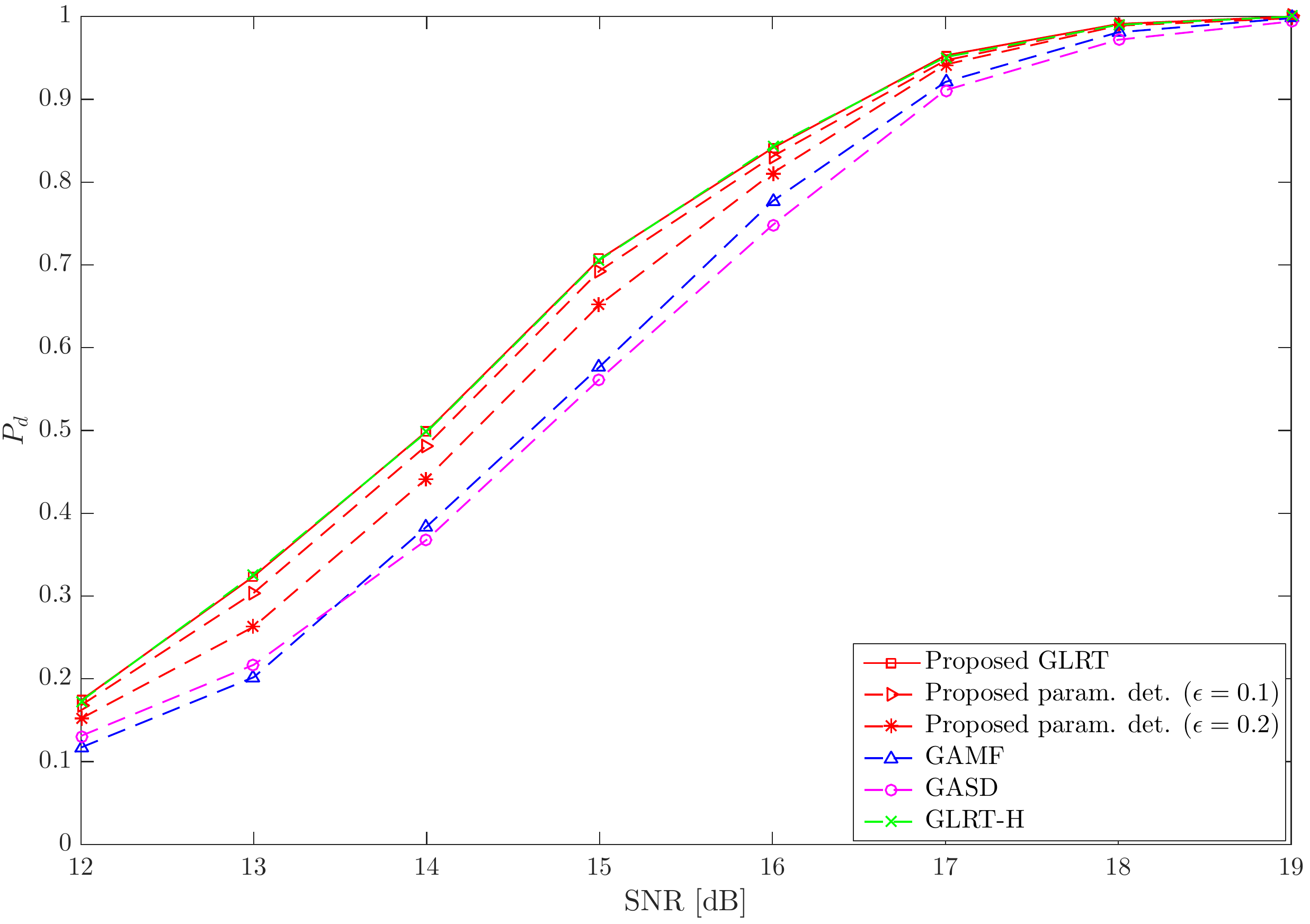}
	\caption{$P_d$ vs SNR under matched conditions, $N=16$, $K_P=4$, and $K_S=40$.}
	\label{fig:matchedKs40}
\end{figure}

\begin{figure}
	\centering
	\includegraphics[width=0.8\textwidth]{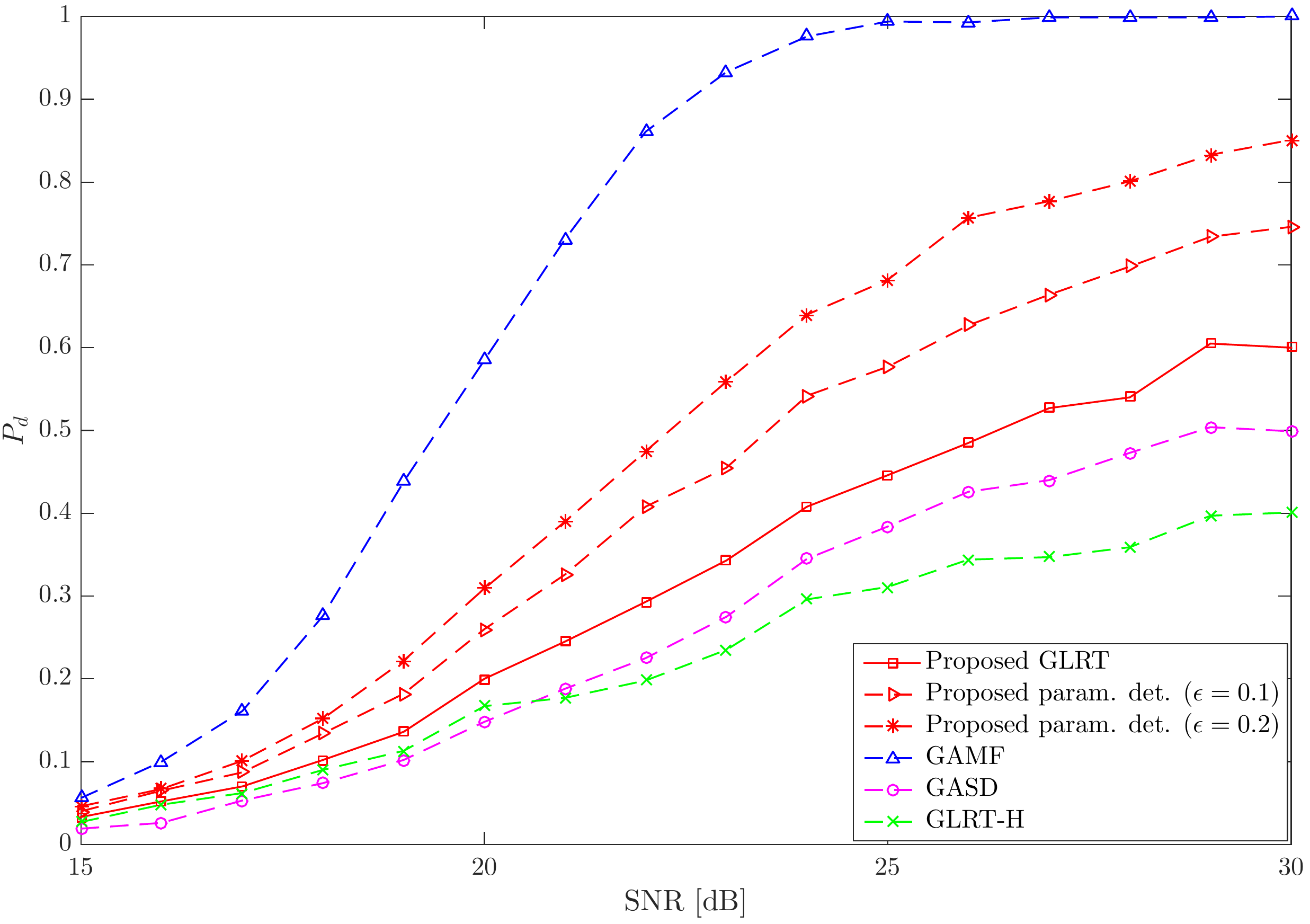}
	\caption{$P_d$ vs SNR in case of mismatched steering vector, for 
	$N=16$, $K_P=4$, $K_S=32$, and
	$\delta = 0.4$ corresponding to $\cos^2 \theta = 0.46$.}
	\label{fig:mismatch04Ks32}
\end{figure}

\begin{figure}
	\centering
	\includegraphics[width=0.8\textwidth]{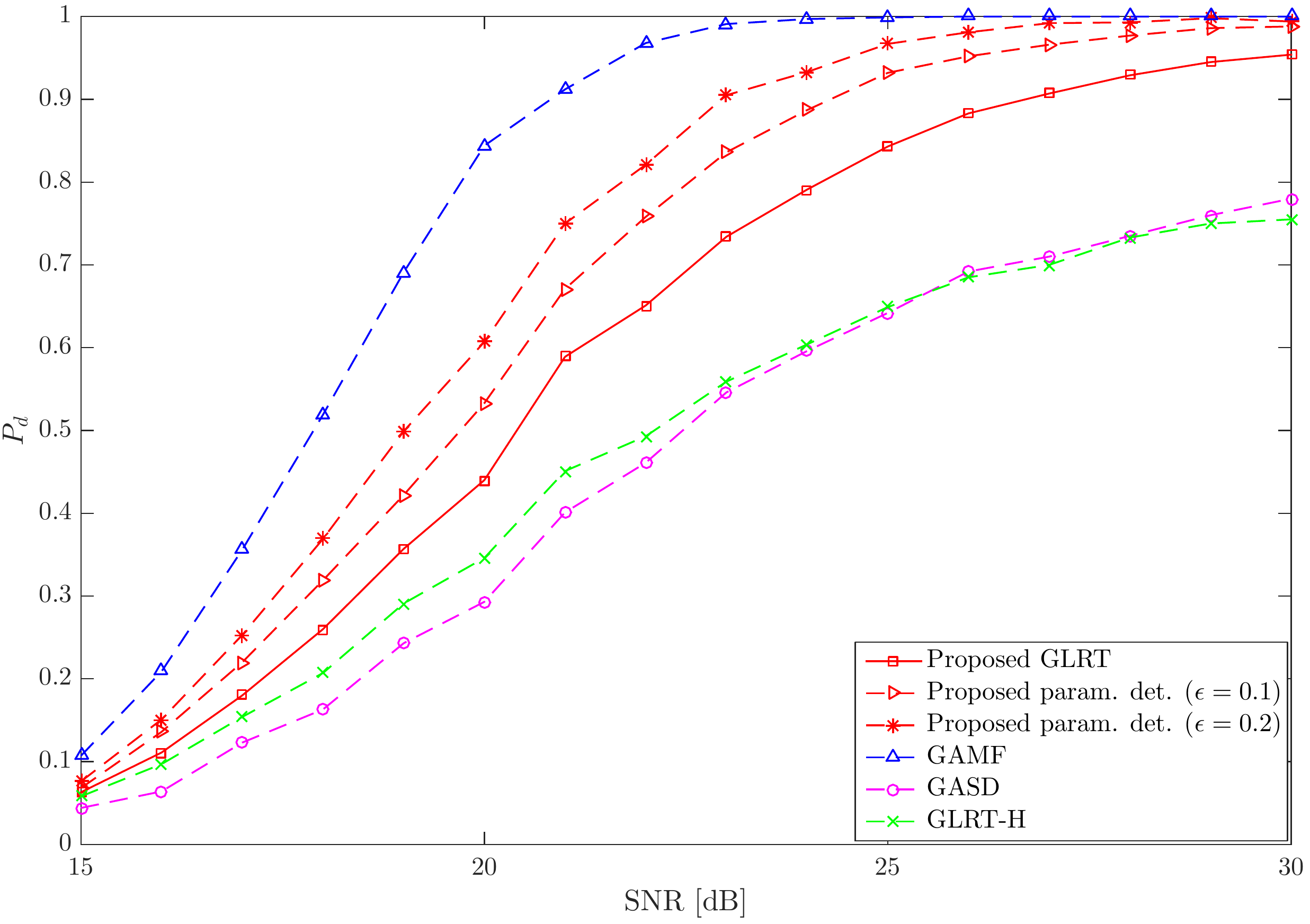}
	\caption{$P_d$ vs SNR in case of mismatched steering vector, for 
	$N=16$, $K_P=4$, $K_S=40$, and
	$\delta = 0.4$ corresponding to $\cos^2 \theta = 0.46$.}
	\label{fig:mismatch04Ks40}
\end{figure}

As first reference scenario, we consider a radar setup with $K_P = 4$ primary data and $K_S = 32$ or $K_S = 40$ training data.
Results under matched conditions are reported in Figs.~\ref{fig:matchedKs32} and \ref{fig:matchedKs40}. 
Figs.~\ref{fig:mismatch04Ks32} and
\ref{fig:mismatch04Ks40} show the results for the mismatched case ($\cos^2 \theta=0.46$). 
For the considered values of $N$, $K_P$, and $K_S$, the analysis shows that, under matched conditions, the GLRT and practically also the parametric detector with $\epsilon=0.1$ guarantee the same performance of the GLRT-H for both $K_S=32,40$. The GAMF and the GASD experience a non-negligible loss with respect to the GLRT-H for both values of secondary data. Finally, the parametric detector with $\epsilon=0.2$ experiences a  very limited loss with respect to the GLRT-H for $K_S=32$. For $K_S=40$ its loss is slightly larger; however, it continues to outperform
the GAMF and the GASD.
Summarizing, under matched conditions, the proposed approach leads to a GLRT that can guarantee the same performance 
of the GLRT-H and to a parametric detector that, depending on the value of its tunable parameter, has the same performance or a limited loss  compared to the GLRT-H. Remarkably, under mismatched conditions, the GLRT and the parametric detector are more robust than the GASD and the GLRT-H. They are less robust than the GAMF; however, the enhanced robustness of the latter is paid  in terms of  a loss for matched signals.

\begin{figure}
	\centering
	\includegraphics[width=0.8\textwidth]{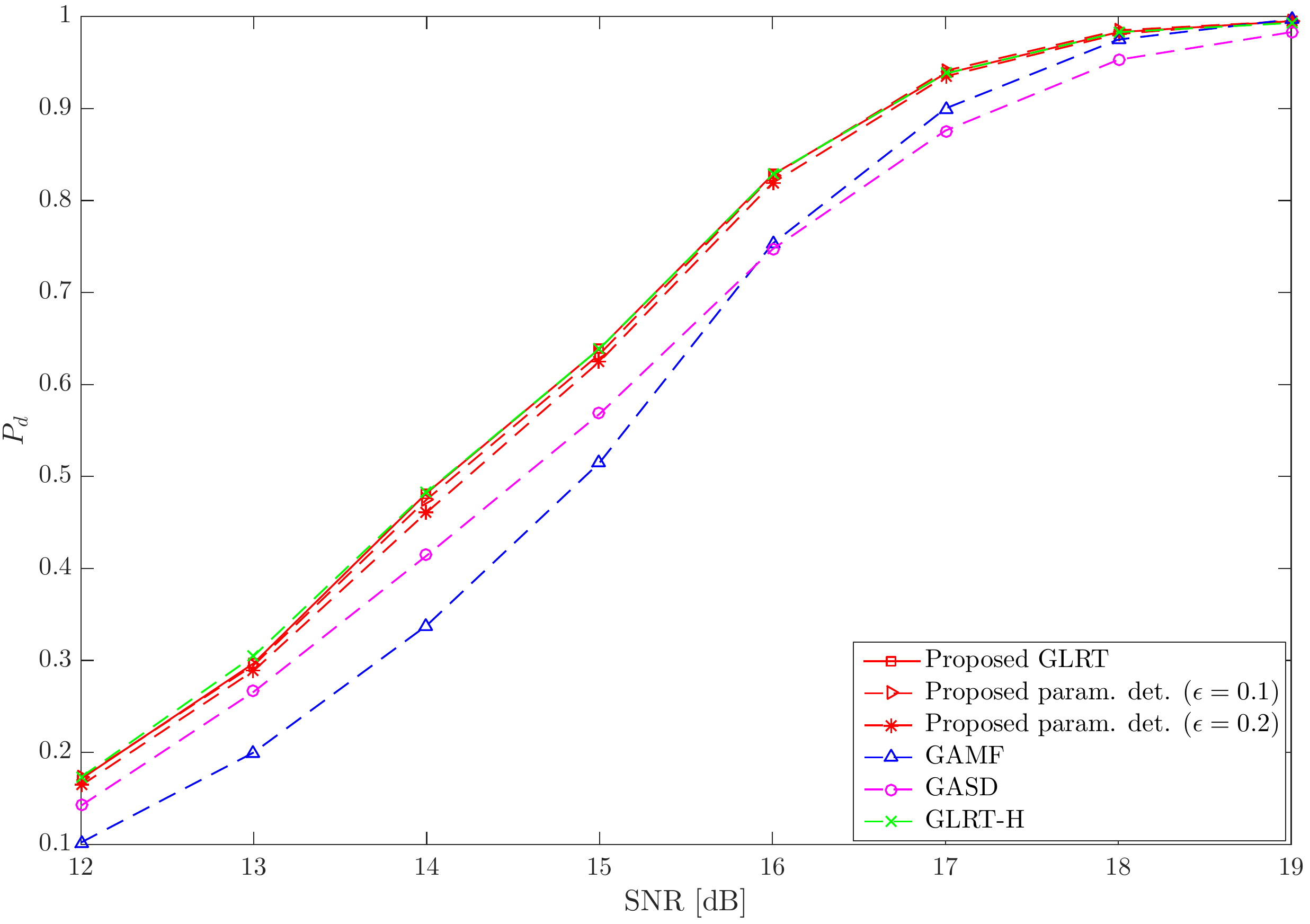}
	\caption{$P_d$ vs SNR under matched conditions, $N=16$, $K_P=2$, and $K_S=32$.}
	\label{fig:matchedKp2Ks32}
\end{figure}

\begin{figure}
	\centering
	\includegraphics[width=0.8\textwidth]{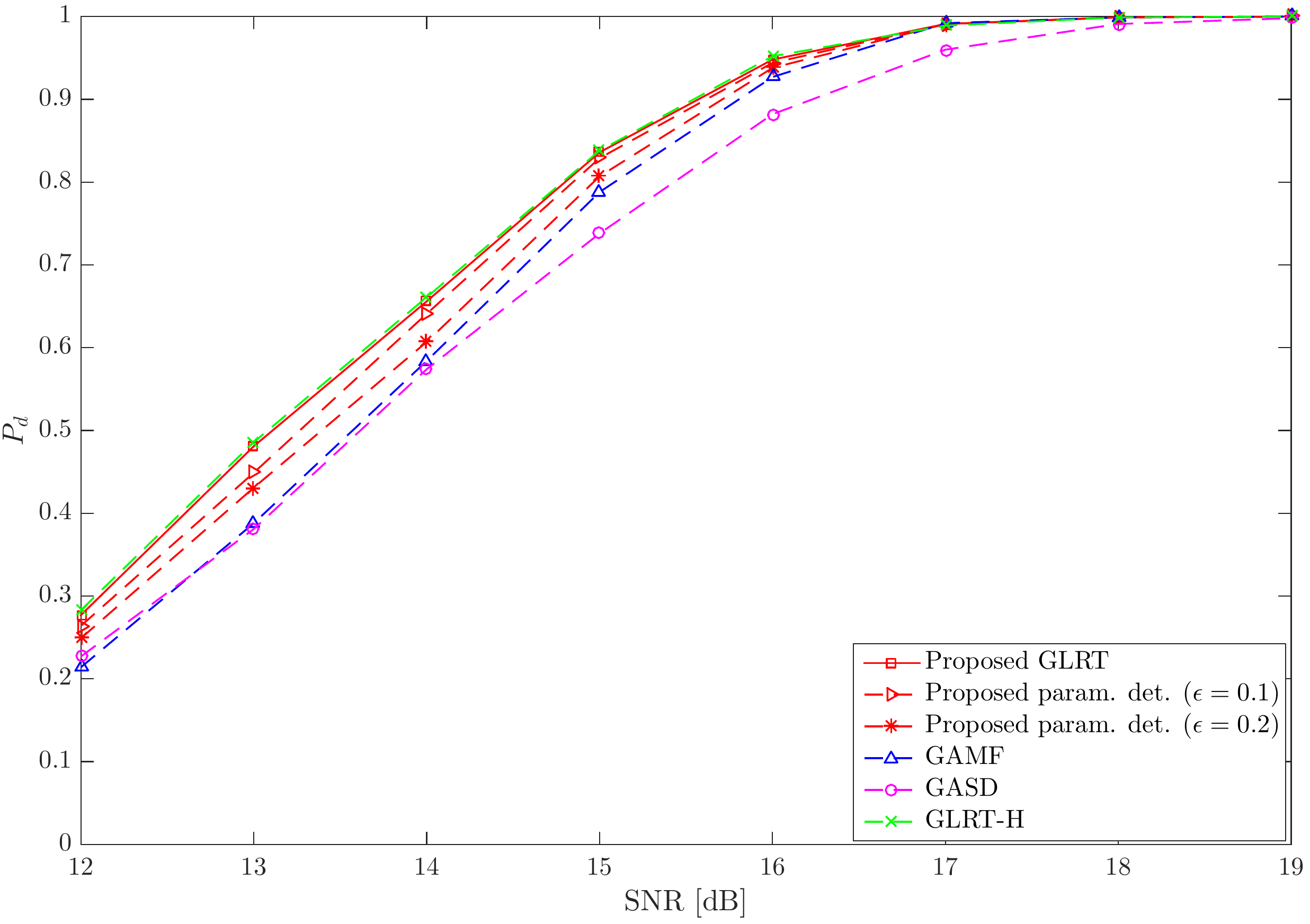}
	\caption{$P_d$ vs SNR under matched conditions, $N=16$, $K_P=2$, and $K_S=40$.}
	\label{fig:matchedKp2Ks40}
\end{figure}

\begin{figure}
	\centering
	\includegraphics[width=0.8\textwidth]{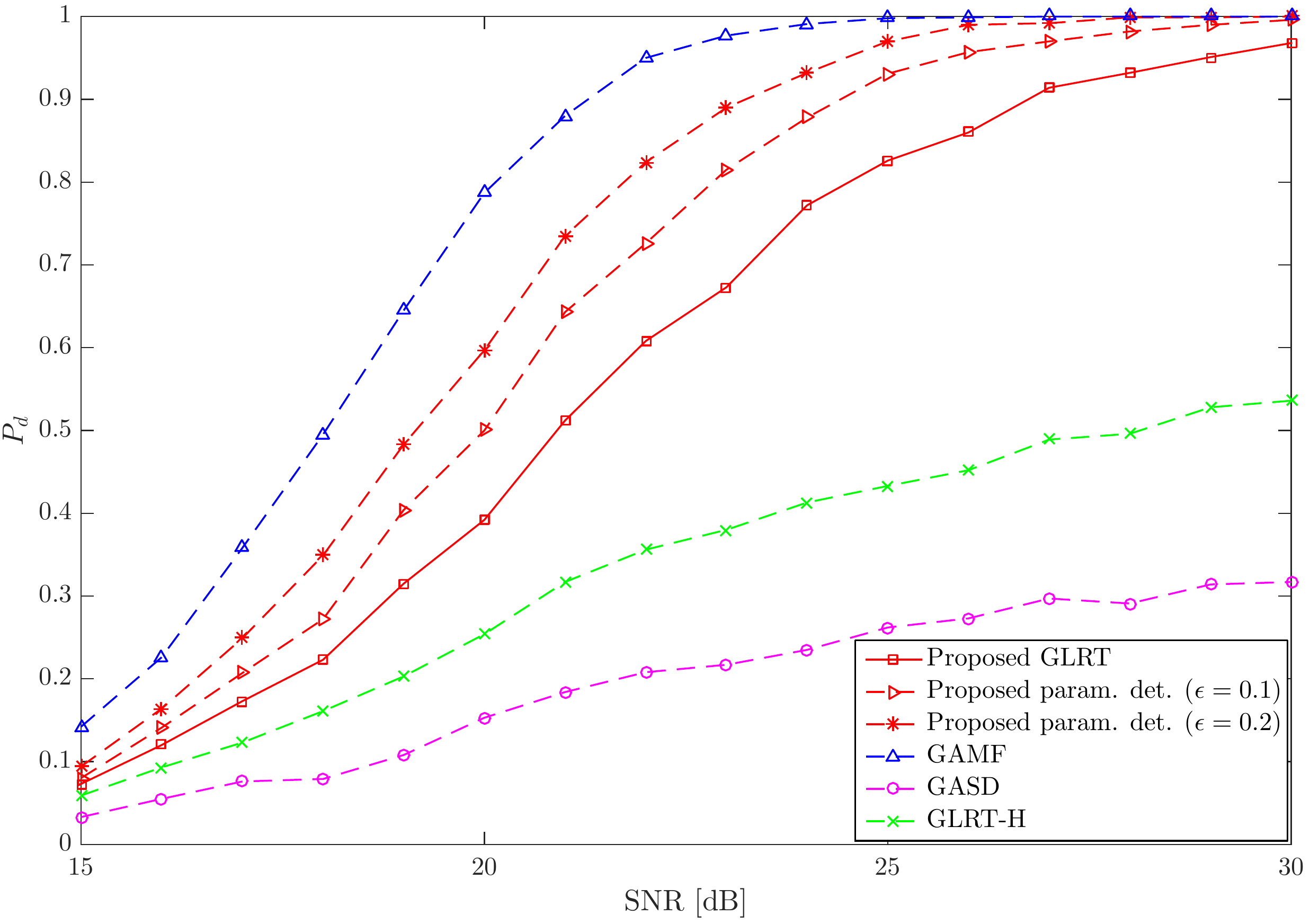}
	\caption{$P_d$ vs SNR in case of mismatched steering vector, for 
	$N=16$, $K_P=2$, $K_S=32$, and
	$\delta = 0.4$ corresponding to $\cos^2 \theta = 0.46$.}
	\label{fig:mismatch04Ks32Kp2}
\end{figure}

\begin{figure}
	\centering
	\includegraphics[width=0.8\textwidth]{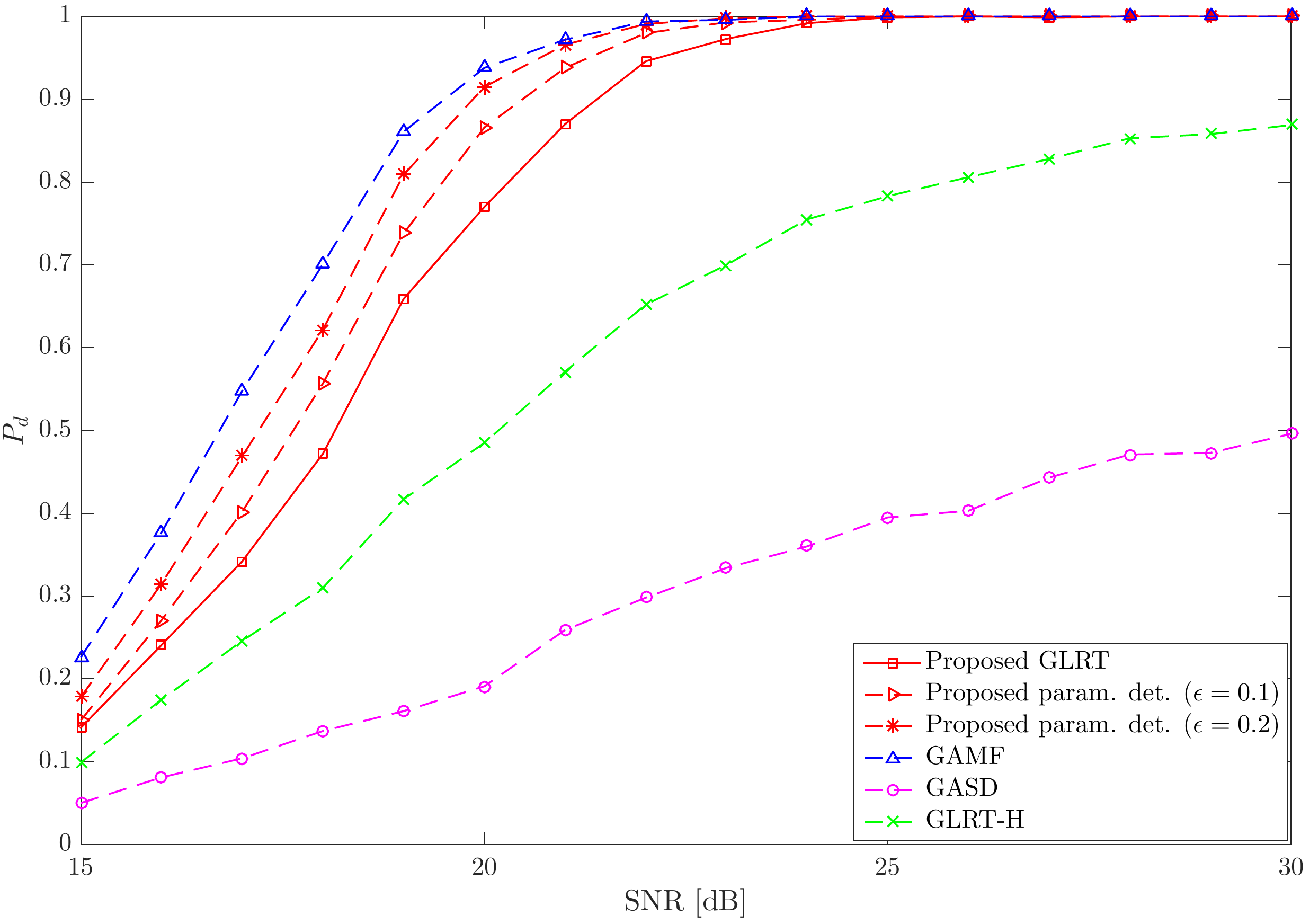}
	\caption{$P_d$ vs SNR in case of mismatched steering vector, for 
	$N=16$, $K_P=2$, $K_S=40$, and
	$\delta = 0.4$ corresponding to $\cos^2 \theta = 0.46$.}
	\label{fig:mismatch04Ks40Kp2}
\end{figure}

A deeper insight into the behavior of the proposed detectors comes from considering additional values of $K_P$. To this end, we also investigate a radar setup assuming $K_P = 2$ and $K_S=32,40$. The results for the matched case
are reported in Figs.~\ref{fig:matchedKp2Ks32}
and \ref{fig:matchedKp2Ks40}
while those for mismatched steering vector ($\cos^2 \theta=0.46$) are given in 
Figs.~\ref{fig:mismatch04Ks32Kp2} and \ref{fig:mismatch04Ks40Kp2} for $K_S = 32$ and $K_S = 40$, respectively. Again,  the loss of the proposed GLRT and of the parametric detector is negligible or limited with respect to the GLRT-H, under matched signals, and they always outperform the GAMF and the GASD. 
As a matter of fact, comparison of Figs.~\ref{fig:mismatch04Ks32} and \ref{fig:mismatch04Ks40} to Figs.~\ref{fig:mismatch04Ks32Kp2} and \ref{fig:mismatch04Ks40Kp2} reveals that the robustness of the proposed detectors increases as the number of primary data $K_P$ decreases. 


\section{Conclusion}

In this paper, we proposed robust CFAR detectors for range-spread targets embedded in Gaussian noise with unknown covariance matrix. The idea is to model the received signal under the signal-plus-noise hypothesis by adding a random component that makes such hypothesis more plausible in presence of mismatches. Moreover, an unknown power of the random component, to be estimated from the observables, limits
the loss with respect to the 
GLRT-H, given  by equation (12) in \cite{Conte-DeMaio-Ricci}, 
when the mismatch is absent. In fact,
the performance assessment shows that the proposed detectors are equivalent or very close to
the GLRT-H, under matched conditions, thus outperforming both the GAMF and the GASD. Under mismatched conditions the proposed detectors are more robust than the 
GLRT-H and the GASD, but  typically less robust than the GAMF whose enhanced robustness is paid in terms of a non-negligble loss under matched conditions. 

\section*{Appendix A: Proof of proposition~\ref{Propo_KellyForsyth}}
\renewcommand{\theequation}{A.\arabic{equation}}


\begin{pf}

Let
$$
m \left( \balpha \right)= 
\det \left(\frac{
\bZ_{\alpha}\bZ_{\alpha}^{\dagger}}{\nu+1}  + \bS \right)
$$
where we recall that
$$
\bZ_{\alpha}= 
\left[ \bz_{1} - \alpha_1  \bv\ \cdots\ \bz_{K_P} - \alpha_{K_P} \bv  \right]
$$
and
start observing that
\begin{eqnarray*}
&&\!\!\!\! \!\!\!\!  \!\!\!\!  \!\!\!\! \!\!\!\! \min_{\balpha \in \C^{K_P}} m \left( \balpha \right)  
\\ &=&
\det (\bS) \min_{\balpha \in \C^{K_P}}  \det \left( \frac{\bS^{-1/2} \bZ_{\alpha} \bZ_{\alpha} ^{\dagger} \bS^{-1/2}}{\nu+1} + \bI_N
\right) \\ &=&
\det (\bS) \min_{\balpha \in \C^{K_P}}  \det \left( \frac{1}{\nu+1} \bZ_{\alpha} ^{\dagger} \bS^{-1} \bZ_{\alpha} 
+ \bI_{K_P}
\right)
\\ &=&
\det (\bS) \min_{\balpha \in \C^{K_P}}  \det \left( \frac{1}{\nu+1} \tilde{\bZ}_{\alpha}^{\dagger} \tilde{\bZ}_{\alpha}
+ \bI_{K_P}
\right)
\end{eqnarray*}
where 
$$
\tilde{\bZ}_{\alpha}= \bS^{-1/2} \bZ_{\alpha}=
\left[ \tilde{\bz}_{1} - \alpha_1  \tilde{\bv} \cdots \tilde{\bz}_{K_P} - \alpha_{K_P} \tilde{\bv}  \right]
$$
with
$\tilde{\bz}_{k} = \bS^{-1/2} \bz_k$ (and $\tilde{\bv} = \bS^{-1/2} \bv$).
It is also easy to check that
$$
\tilde{\bZ}^{\dagger}_{\alpha} \tilde{\bZ}_{\alpha} = \tilde{\bZ}^{\parallel \dagger}_{\alpha} \tilde{\bZ}^{\parallel}_{\alpha} + \tilde{\bZ}^{\perp \dagger} \tilde{\bZ}^{\perp}
$$
where
\begin{eqnarray*}
\tilde{\bZ}^{\parallel}_{\alpha} &=& \Pvtilde \tilde{\bZ}_{\alpha} = \Pvtilde \left[ \tilde{\bz}_{1} - \alpha_1  \tilde{\bv} \ \cdots \ \tilde{\bz}_{K_P} - \alpha_{K_P} \tilde{\bv}  \right]
\\ &=& \left[ \Pvtilde \tilde{\bz}_{1} - \alpha_1  \tilde{\bv} \ \cdots \ \Pvtilde \tilde{\bz}_{K_P} - \alpha_{K_P} \tilde{\bv}  \right]
\end{eqnarray*}
and
\begin{eqnarray*}
\tilde{\bZ}^{\perp}
&=& \Pvtildeperp \tilde{\bZ}_{\alpha} = \Pvtildeperp \left[ \tilde{\bz}_{1} - \alpha_1  \tilde{\bv} \ \cdots \ \tilde{\bz}_{K_P} - \alpha_{K_P} \tilde{\bv}  \right]
\\ &=& \left[ \Pvtildeperp \tilde{\bz}_{1} \ \cdots \ \Pvtildeperp \tilde{\bz}_{K_P} \right]
\end{eqnarray*}
with $\Pvtilde=\bS^{-1/2} \bv \left( \bv^{\dagger} \bS^{-1} \bv \right)^{-1} \bv^{\dagger} \bS^{-1/2}$ and $\Pvtildeperp = \bI_N-\Pvtilde$ the projection matrices onto 
the space spanned by $\bvtilde$ and its orthogonal complement, respectively.
Thus, we have that
\begin{eqnarray*}
m_{\min} &=& \det (\bS)
\min_{\balpha \in \C^{K_P}}  \det \left( \frac{1}{\nu+1} \tilde{\bZ}^{\dagger}_{\alpha} \tilde{\bZ}_{\alpha}
+ \bI_{K_P}
\right)
\\ &=& 
\det (\bS) \min_{\balpha \in \C^{K_P}}  \det \left[ \frac{\tilde{\bZ}^{\parallel \dagger}_{\alpha} \tilde{\bZ}^{\parallel}_{\alpha} + \tilde{\bZ}^{\perp \dagger} \tilde{\bZ}^{\perp}}{\nu+1} + \bI_{K_P}
\right]
\\ &=&
\det (\bS) \det \left[ \frac{1}{\nu+1}  \tilde{\bZ}^{\perp \dagger} \tilde{\bZ}^{\perp} + \bI_{K_P} \right]
\end{eqnarray*}
where we used the inequality \cite{handbook_of_matrices}
$$
\det \left(\bA +\bB \right) \geq \det \left( \bA \right)
$$
that is satisfied by any choice of the Hermitian and positive matrix $\bA \in \C^{N \times N}$ and the Hermitian and positive semidefinite matrix $\bB\in \C^{N \times N}$. As a matter of fact, $\tilde{\bZ}_{\alpha}^{\parallel}$ is a matrix of zeroes if
$\Pvtilde \tilde{\bz}_{k} - \alpha_k  \tilde{\bv}=\bzero$, $k=1, \ldots, K_P$,
that leads to
\begin{eqnarray*}
\widehat{\alpha}_k &=& \frac{ \tilde{\bv}^{\dagger} \Pvtilde \tilde{\bz}_{k}}{\tilde{\bv}^{\dagger} \tilde{\bv}}=
\frac{\tilde{\bv}^{\dagger} \Pvtilde \bS^{-1/2} \bz_{k}}{\bv^{\dagger} \bS^{-1} \bv}
\\ &=&
\frac{ \bv^{\dagger} \bS^{-1} \bv (\bv^{\dagger} \bS^{-1} \bv)^{-1} \bv^{\dagger} \bS^{-1/2} \bS^{-1/2} \bz_{k}}{\bv^{\dagger} \bS^{-1} \bv}
\\ &=&
\frac{  \bv^{\dagger} \bS^{-1} \bz_{k}}{\bv^{\dagger} \bS^{-1} \bv}, \qquad k=1, \ldots, K_P.
\end{eqnarray*}
Obviously, the minimum can be re-written as
\begin{eqnarray*}
m_{\min} &=&
\det (\bS) \det \left[ \frac{1}{\nu+1}   \tilde{\bZ}^{\perp} \tilde{\bZ}^{\perp \dagger} + \bI_{N} \right]
\\ &=&
\det (\bS) \det \left[ \frac{\sum_{k=1}^{K_P} \Pvtildeperp \tilde{\bz}_k \tilde{\bz_k}^{\dagger}  \Pvtildeperp}{\nu+1} + \bI_{N} \right]
\\ &=&
\det (\bS) \det \left[ \frac{\Pvtildeperp \bS^{-1/2} \bZ \bZ^{\dagger} \bS^{-1/2} \Pvtildeperp}{\nu+1} + \bI_{N} \right]
\end{eqnarray*}
but also as
\begin{eqnarray*}
m_{\min} &=&
\det (\bS) \det \left[ \frac{1}{\nu+1}  \tilde{\bZ}^{\perp \dagger} \tilde{\bZ}^{\perp} + \bI_{K_P} \right]
\\ &=&
\det (\bS) \det \left[ \frac{1}{\nu+1}  \bZ_{\widehat{\alpha}}^{\dagger} \bS^{-1/2} \bS^{-1/2} \bZ_{\widehat{\alpha}}+ \bI_{K_P} \right]
\\ &=&
\det (\bS) \det \left[ \frac{1}{\nu+1} \bS^{-1/2} \bZ_{\widehat{\alpha}}  \bZ_{\widehat{\alpha}}^{\dagger} \bS^{-1/2}
+ \bI_{N} \right]
\\ &=&
\det \left[ \frac{1}{\nu+1} \bZ_{\widehat{\alpha}}  \bZ_{\widehat{\alpha}}^{\dagger} 
+ \bS \right]
\end{eqnarray*}
with
$$
\bZ_{\widehat{\alpha}}= \left[ \bz_{1} - \widehat{\alpha}_1 \bv \ \cdots \ \bz_{K_P} - \widehat{\alpha}_{K_P} \bv  \right].
$$
\hfill$\square$
\end{pf}


\begin{thebibliography}{99}


\bibitem{Kelly_techrep}
E. J. Kelly and K. Forsythe, ``Adaptive Detection and Parameter
Estimation for Multidimensional Signal Models,'' Lincoln Laboratory, MIT,
Lexington, MA, Tech. Rep. No. 848, Apr. 19, 1989.

\bibitem{CAI-WANG}
H. Wang and L. Cai,
``On Adaptive Multiband Signal Detection with GLR Algorithm,''
{\em IEEE Trans. Aerosp. Electron. Syst.}, Vol. 27, No.~2, pp.~225-233, Mar.~1991.


\bibitem{Conte-DeMaio-Ricci}
E. Conte, A.~De Maio, and G. Ricci,
``GLRT-based Adaptive Detection Algorithms for Range-Spread Targets,''
{\em IEEE Trans. Signal Process.},
Vol.~49, No.~7, pp.1336-1348, Jul. 2001.


\bibitem{PartialObs}
L. Xiao, Y. Liu, T. Huang, X. Liu, and  X. Wang,
``Distributed Target Detection with Partial Observation,''
{\em IEEE Trans. Signal Process.},
Vol.~66, No.~6, pp.1551-1565, 15 Mar. 2018.

\bibitem{Ciunzo}
D. Ciuonzo, A. De Maio, and D. Orlando, ``A Unifying Framework for Adaptive Radar Detection in Homogeneous plus Structured Interference-Part II:
Detectors Design,'' {\em IEEE Trans. Signal Process.}, Vol. 64, No. 11, pp. 2907-2919, Jun. 2016.


\bibitem{Lombardo}
P. Lombardo and D. Pastina, 
``Multiband coherent radar detection against compound-Gaussian clutter,''
{\em IEEE Trans. Aerosp. Electron. Syst.}, Vol. 35, No.~4, pp.~1266-1282, Oct.~1999.

\bibitem{ConteDeMaio}
E. Conte and A. De Maio,
``Distributed target detection in compound-Gaussian noise with Rao and Wald tests,''
{\em IEEE Trans. Aerosp. Electron. Syst.}, Vol. 39, No.~2, pp.~568-582, Apr.~2003.

\bibitem{Guan2011}
J. Guan and X. Zhang,
``Subspace detection for range and Doppler distributed targets with
Rao and Wald tests,'' {\em Signal Process.}, Vol. 91, pp. 51-60, 2011. 



%
%

\bibitem{BOR-MorganClaypool}
F. Bandiera, D. Orlando, and G. Ricci,
``Advanced Radar Detection Schemes Under Mismatched Signal Models,''
{\em Synthesis Lectures on Signal Processing No. 8, Morgan \& Claypool Publishers},
2009.

\bibitem{Bandiera-Orlando-Ricci-conidistribuiti}
F. Bandiera, D. Orlando, and G. Ricci, 
``CFAR Detection strategies for Distributed Targets under Conic Constraints,''
{\em IEEE Trans. Signal Process.},
Vol. 57, No. 9, pp. 3305-3316, Sept. 2009.

\bibitem{Pulsone}
N. B. Pulsone and C. M. Rader, ``Adaptive Beamformer Orthogonal Rejection Test,'' {\em IEEE Trans. Signal Process.},
Vol. 49, No. 3, pp. 521-529, Mar. 2001.


\bibitem{BBR_WA}
F. Bandiera, O. Besson, and G. Ricci, 
``An ABORT-like Detector with Improved Mismatched Signals Rejection Capabilities'',
{\em IEEE Trans. Signal Process.},
Vol.~56, No.~1, pp.~ 14-25, Jan. 2008.

\bibitem{Orlando}
C. Hao, J. Yang, X. Ma, C. Hou, and D. Orlando,
``Adaptive detection of distributed targets with orthogonal rejection,''
{\em IET Radar, Sonar \& Navigation}, Vol. 6, No. 6, pp. 483-493, Jul. 2012. 

\bibitem{Friedlander}
Y. Jin and B. Friedlander, ``A CFAR Adaptive Subspace Detector for Second-Order Gaussian Signals,'' {\em IEEE Trans. Signal Process.}, Vol. 53, No. 3, pp. 871-884, Mar. 2005.

\bibitem{Ricci}
G. Ricci and L. L. Scharf, ``Adaptive Radar Detection of Extended Gaussian Targets,'' {\em Proc. 12th Annual Workshop on Adaptive Sensor Array Processing}, Lincoln Laboratory, MIT, Lexington, Massachusetts (USA), 16-18 Mar. 2004.  

\bibitem{CAMSAP}
O. Besson, E. Chaumette, and F. Vincent,
``Adaptive Detection of a Gaussian Signal in Gaussian Noise,''
{\em Proc. 6th IEEE Int. Workshop Comput. Adv. in Multi-Sens. Adaptive 
Process.}, Canc\'un, Mexico, Dec. 13-16, 2015, pp. 117-120.

\bibitem{Besson_collaboration}
O. Besson, A. Coluccia, E. Chaumette, G. Ricci, and F. Vincent,
``Generalized likelihood ratio test for detection of Gaussian rank-one signals 
in Gaussian noise with unknown statistics,''
{\em IEEE Trans. Signal Process.}, Vol. 65, No. 4, pp. 1082-1092, 15 Feb. 2017.


\bibitem{CISS2018}
A. Coluccia and G.~Ricci, 
``A random-signal approach to robust radar detection,''
{\em Proc. 52nd Annual Conference on Information Sciences and Systems (CISS)},
Princeton, NJ, USA, 21-23 Mar. 2018.


\bibitem{BCRSigProc2018}
A. Coluccia, G. Ricci, and O. Besson,
``Design of robust radar detectors through random perturbation of the target signature,''
http://arxiv.org/abs/1903.08468 (also submitted to {\em IEEE Trans. on Signal Process.}).

\bibitem{Kelly}
E. J. Kelly, ``An Adaptive Detection Algorithm,'' {\em IEEE Trans. Aerosp. Electron. Syst.}, Vol. 22, No.~2, pp.~115-127, Mar.~1986.

\bibitem{Magnus}
J. R. Magnus, H. Neudecker,
{\em Matrix Differential Calculus with Applications in Statistics and Econometrics},
John Wiley \& Sons, 1999.

\bibitem{Tourneret}
J.-Y. Tourneret, A. Ferrari, and G. Letac,
``The Noncentral Wishart distribution: properties and application to speckle imaging,''
{\em Proc. 13th Workshop on Statistical Signal Processing}, Bordeaux, France, 17-20 Jul. 2005.

\bibitem{McKay}
M. R. McKay and I. B. Collings,
``Statistical Properties of Complex Noncentral Wishart Matrices and MIMO Capacity,''
{\em Proc. International Symposium on Information Theory}, Adelaide, Australia, 4-9 Sept. 2005, pp. 785-789.


\bibitem{Kelly-Nitzberg}
F.~C.~Robey, D.~L.~Fuhrman, E.~J.~Kelly, and R.~Nitzberg,
``A CFAR Adaptive Matched Filter Detector,''
{\em IEEE Trans. Aerosp. and Electron. Syst.},
Vol.~29, No.~1, pp.~208-216, Jan.~1992.


\bibitem{Asymptotically}
E. Conte, M. Lops, and G. Ricci, ``Asymptotically Optimum Radar Detection
in Compound Gaussian Noise,'' {\em IEEE Trans. Aerosp. and Electron. Syst.},
Vol.~31, No. 2, pp.~617-625, April~1995.

\bibitem{ACE}
S. Kraut and L. L. Scharf, ``The CFAR adaptive subspace detector is
a scale-invariant GLRT,'' {\em IEEE Trans. Signal Process.}, Vol.~47, No. 9, pp.~2538-2541, Sept. 1999.

\bibitem{Gini}
F. Gini,
``Sub-optimum coherent radar detection in a mixture of
K-distributed and Gaussian clutter,'' {\em IEE Proceedings - Radar, Sonar and Navigation},
Vol. 144, No. 1, pp. 39-48, Feb. 1997.

\bibitem{handbook_of_matrices}
H. L\"{u}tkepohl, {\em Handbook of Matrices}, John Wiley \& Sons, 1996.

\end{thebibliography}
\end{document}